\documentclass[twocolumn,prb,aps,superscriptaddress]{revtex4-1}
\usepackage[latin9]{inputenc}
\usepackage{mathtools}
\usepackage{amsbsy}
\usepackage{amstext}
\usepackage[unicode=true,pdfusetitle,
 bookmarks=false,
 breaklinks=false,pdfborder={0 0 1},backref=false,colorlinks=false]
 {hyperref}
\hypersetup{
 bookmarksnumbered=false,bookmarksopen=false}
\makeatletter

\@ifundefined{textcolor}{}
{%
 \definecolor{BLACK}{gray}{0}
 \definecolor{WHITE}{gray}{1}
 \definecolor{RED}{rgb}{1,0,0}
 \definecolor{GREEN}{rgb}{0,1,0}
 \definecolor{BLUE}{rgb}{0,0,1}
 \definecolor{CYAN}{cmyk}{1,0,0,0}
 \definecolor{MAGENTA}{cmyk}{0,1,0,0}
 \definecolor{YELLOW}{cmyk}{0,0,1,0}
}

\usepackage{amsmath}
\usepackage{graphicx}
\usepackage{amssymb}
\usepackage{txfonts,color}
\makeatother

\begin{document}
\title{Selective Hybrid Spin Interactions with Low Radiation Power}
\author{I. Arrazola}
\affiliation{Department of Physical Chemistry, University of the Basque Country UPV/EHU, Apartado 644, 48080 Bilbao, Spain}
\author{E. Solano}
\affiliation{Department of Physical Chemistry, University of the Basque Country UPV/EHU, Apartado 644, 48080 Bilbao, Spain}
\affiliation{IKERBASQUE,  Basque  Foundation  for  Science,  Maria  Diaz  de  Haro  3,  48013  Bilbao,  Spain}
\affiliation{Department of Physics, Shanghai University, 200444 Shanghai, China}
\author{J. Casanova}
\affiliation{Department of Physical Chemistry, University of the Basque Country UPV/EHU, Apartado 644, 48080 Bilbao, Spain}
\affiliation{IKERBASQUE,  Basque  Foundation  for  Science,  Maria  Diaz  de  Haro  3,  48013  Bilbao,  Spain}

\begin{abstract}
We  present a protocol for designing appropriately extended $\pi$ pulses that achieves tunable, thus selective, electron-nuclear spin interactions with low-driving radiation power. The latter is of great benefit when $\pi$ pulses are displayed over biological samples as it reduces sample heating. Our method is general since it can be applied to different quantum sensor devices such as nitrogen vacancy centers or silicon vacancy centers. Furthermore, it can be directly incorporated in commonly used stroboscopic dynamical decoupling techniques to achieve enhanced nuclear selectivity and control, which demonstrates its flexibility. 
\end{abstract}
\maketitle

\section{Introduction}
Nanoscale nuclear magnetic resonance (NMR) emerged as a promising research field~\cite{Degen17}, with the priority goal of detecting and controlling  magnetic field emitters (as nuclear spins) with high frequency and spatial resolution~\cite{Rondin14, Schmitt17, Boss17, Zopes18, Glenn18, Zopes18bis}. This is achieved with the help of a quantum sensor that appears, e.g., when a diamond is doped with different impurities resulting in an optically active diamond sample~\cite{Walker79}. Consequently, these impurities  receive the name of color centers~\cite{Aharonovich11}. Among frequent color centers one can find in diamond, we can mention, e.g., the nitrogen vacancy (NV) center~\cite{Doherty13, Suter16}, or the silicon vacancy center~\cite{Rogers14}.  These  carry an electronic spin that allows fast external control with microwave (MW) radiation, while they can be initialised and measured with optical fields~\cite{Schirhagl14, Wu16}. In particular, the NV center is a prominent quantum sensor candidate owing to its long decay time (or longitudinal relaxation time) of the order of milliseconds at room temperature~\cite{Degen17}. Furthermore, at low temperatures of  $\approx 3.7$~K longitudinal relaxation times approaching to $10^3$ s have been recently reported~\cite{Abobeih18}. Another error source is that affecting the quantum coherence of the sensor. This mainly appears as a consequence of the interaction among $^{13}$C nuclei in the diamond and the NV~\cite{Maze08}. However, with the help of dynamical decoupling (DD) techniques~\cite{Souza12}, one can efficiently remove this error source and take the coherence time $T_2$ to the decay time~$T_1$~\cite{Souza12}.

From a different perspective, DD techniques are also employed to couple the NV to a target signal. The latter being classical electromagnetic radiation~\cite{Taylor08}, or the hyperfine fields emitted by nuclear spins~\cite{Abobeih18, Muller14, Wang16}. In particular, DD techniques generate filters that allow the passage of signals with only specific frequencies~\cite{Hasse18}.  It is the accuracy of this filter what determines the fidelity in detection and control on the target signal.  Continuous and pulsed (or stroboscopic) DD schemes are typically considered. While the former requires to fulfill the Hartmann-Hahn condition~\cite{Hartmann62, Casanova18bis}, the latter uses the time spacing among $\pi$ pulses to induce a rotation frequency in the NV matching that of the target signal~\cite{Taminiau12}. Pulsed DD schemes have advantages over continuous DD methods such as the achievement of enhanced frequency selectivity by using large harmonics of the generated modulation function~\cite{Taminiau12, Casanova15}. Another advantage is the demonstrated robustness against control errors of certain pulse sequences such as those of the XY family~\cite{Maudsley86, Gullion90, Souza11, Wang17, Arrazola18}. However, the use of large harmonics makes DD sequences sensitive to environmental noise, and leads to signal overlaps which hinders  spectral readout~\cite{Casanova15}. As we will show, these issues can be minimised by applying a large static magnetic field $B_z$.  Unfortunately, the performance of pulsed DD techniques under large $B_z$ gets spoiled unless $\pi$ pulses are fast, i.e. highly energetic, compared with nuclear Larmor frequencies (note these are proportional to $B_z$). This represents a serious disadvantage, especially when DD sequences act over biological samples, since fast $\pi$ pulses require high MW power causing damage as a result of the induced heating~\cite{Cao17}.

In this article, we propose a design of amplitude modulated decoupling pulses that solves these problems and achieves tunable, hence highly selective, NV-nuclei interactions. This can be done without fast $\pi$ pulses, i.e. with low MW power, and involving large magnetic fields. We use an NV center in diamond to illustrate our method, although this is general thus applicable to arbitrary hybrid spin systems. Furthermore, our protocol can be incorporated to standard pulsed DD sequences such as the widely used XY-8 sequence, demonstrating its flexibility. We note that a different approach based on a specific continuous DD method~\cite{Casanova18bis} has been proposed to operate with NV centers under large $B_z$ fields.

\section{Model}
We consider an NV center coupled to nuclear spins and under an external MW driving. This is described by 
\begin{equation}\label{original}
H = D S_z^2 - \gamma_e B_z S_z -\sum_j \omega_{\rm L} I_j^z + S_z \sum_j \vec{A}_j \cdot \vec{I}_j + H_{\rm c},
\end{equation}
where $D = (2\pi)\times 2.87$ GHz, $\gamma_e = -(2\pi)\times 28.024$ GHz/T  is the electronic gyromagnetic ratio, and $B_z$ is applied in the NV axis (the $z$ axis). The nuclear Larmor frequency  $\omega_{\rm L} = \gamma_n B_z$ with $\gamma_n$ the nuclear gyromagnetic ratio.  $S_z =|1\rangle\langle 1|-|-\!1\rangle\langle -1| $ with $|1\rangle$ and  $|\!-\!\!1\rangle$ the hyperfine levels of the NV. The nuclear spin-$1/2$ operators $I_j^\alpha = 1/2 \ \sigma_j^{\alpha}$ ($\alpha = x, y ,z$) and  $\vec{A}_j$ is the hyperfine vector mediating NV-nucleus coupling. The control Hamiltonian  $H_{\rm c} = \sqrt{2} \Omega(t) S_x \cos{[\omega t -\phi]}$  ($\phi$ is the pulse phase) with 
 $S_x = \frac{1}{\sqrt 2} (|1\rangle\langle 0| +|\!-\!1\rangle\langle 0| + {\rm H.c.})$, and $\omega$ is the MW driving frequency on resonance with the $|1\rangle \leftrightarrow |0\rangle$ NV transition. In the rotating frame of $D S_z^2 - \gamma_e B_z S_z$, Eq.~(\ref{original}) reads 
\begin{equation}\label{simulations}
H = \sum_j \omega_j \ \hat{\omega}_j\cdot \vec{I}_j + \frac{\sigma_z}{2}\sum_j \vec{A}_j\cdot \vec{I}_j + \frac{\Omega(t)}{2} (|1\rangle\langle 0| e^{i\phi} + {\rm H.c.}).
\end{equation}
The $j$th nuclear resonance frequency is $\omega_j \approx \omega_{\rm L}- \frac{1}{2}A_j^z$, and $\hat{\omega}_j = \vec{\omega}_j / |\vec{\omega}_j|$ with $\vec{\omega}_j = \omega_{\rm L} \hat{z} - \frac{1}{2} \vec{A}_j$ (note that $|\vec{\omega}_j| = \omega_j$). Furthermore, we call $H'_{\rm c} = \frac{\Omega(t)}{2} (|1\rangle\langle 0| e^{i\phi} + {\rm H.c.})$. 

Pulsed DD methods rely on the stroboscopic application of the MW driving (i.e. of $H'_{\rm c}$) leading to periodic $\pi$ rotations in the NV electronic spin. This is described by the  effective Hamiltonian (in the rotating frame of $H'_{\rm c}$)
$H =  -\sum_j \omega_j \ \hat{\omega}_j\cdot \vec{I}_j +F(t) \frac{\sigma_z}{2}\sum_j \vec{A}_j\cdot \vec{I}_j$, with the modulation function $F(t)$ taking periodically the values $+1$ or $-1$, depending on the number of $\pi$ pulses on the NV. 

A common assumption of standard DD techniques is that  $\pi$ pulses are nearly instantaneous, thus highly energetic. However, in real cases we deal with finite-width pulses such that, e.g., when caused by a $H'_{\rm c}$ with constant $\Omega$, a time $t_{\pi} = \frac{\pi}{\Omega}$ is needed to produce a $\pi$ pulse. This has adverse consequences on the NV-nuclei dynamics such as the appearance of spurious resonances~\cite{Loretz15, Haase16, Lang17}, or the drastic reduction of the  NMR sensitivity at large $B_z$~\cite{Casanova18}.  Note that, in Ref.~\cite{Casanova18} a strategy to signal recovery is also  presented, while that approach does not lead to selective nuclear interactions. However, we will demonstrate that the introduction of extended pulses with tailored $\Omega$ leads to tunable NV-nuclei interactions with low power MW radiation. 

\section{Dynamical decoupling with instantaneous pulses}
\begin{figure}[t]
\hspace{-0.0 cm}\includegraphics[width=0.85\columnwidth]{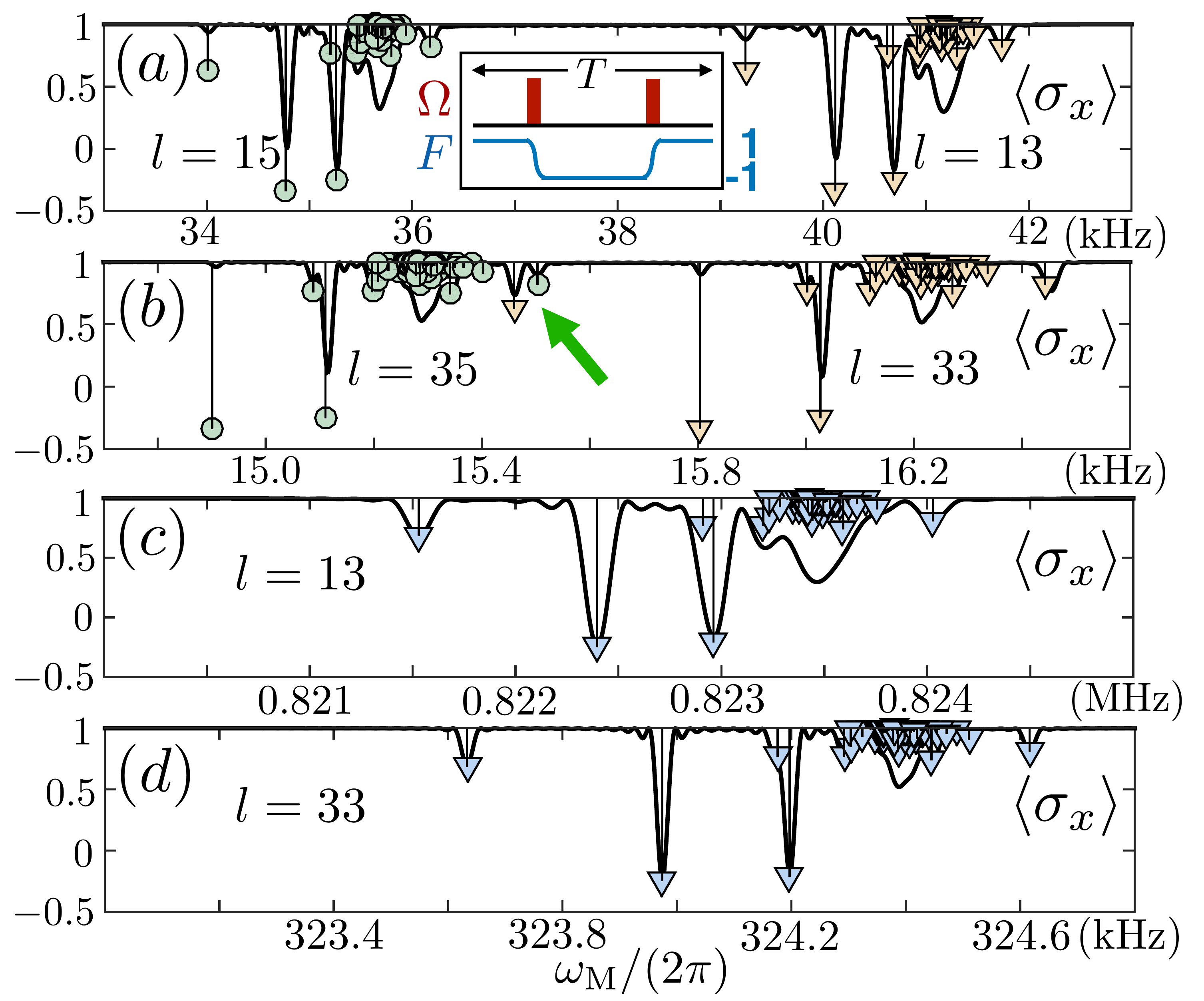}
\caption{Signal  (black-solid)  harvested with instantaneous $\pi$ pulses, $ B_z = 500$ G in (a) (b) and $B_z = 1$ T in (c) (d). Circles and triangles are the theoretically expected values for $\langle \sigma_x \rangle$. In (a) we select $l=13, 15$ and their signals are clearly separated. In (b) we use $l=33, 35$ and observe a spectral overlap (green arrow). In (c) (d) the spectral overlap is removed owing to a large $B_z$, while the signal (black-solid) matches the theoretically expected values. Final sequence time for (a) (c) is $\approx 0.5$ ms, and $\approx 1.2$ ms for (b) (d).}
\label{instantaneous}
\end{figure}

We consider the widely used XY-8=XYXYYXYX scheme, with X (Y) a $\pi$ pulse over the $x$ ($y$) axis. The  sequential application of XY-8 on the NV leads to a periodic, even, $F(t)$ that  expans in harmonic functions as $F(t)=\sum_n f_n \cos{(n\omega_{\rm M} t)}$, where $f_n = 2/T\int_0^T F(s)  \cos{(n \omega_{\rm M} s)} ds$, and $\omega_{\rm M} = \frac{2\pi}{T}$  with $T$ the period of $F(t)$. See an example of $F(t)$ in the inset of Fig.~\ref{instantaneous} (a). In the rotating frame of $-\sum_j \omega_j \ \hat{\omega}_j\cdot \vec{I}_j $, Eq.~(\ref{simulations}) is 
\begin{equation}\label{modulated}
H = \sum_{n,j} \frac{f_n \cos{(n \omega_{\rm M} t)} \sigma_z}{2} \bigg[A_{j}^{x} I^x_j   \cos{(\omega_j t)} + A_{j}^{y} I^y_j   \sin{(\omega_j t)} + A_{j}^{z} I^z_j\bigg],
\end{equation}
where $A_{j}^{x,y} = |\vec{A}_{j}^{x,y}|$ with $\vec{A}_{j}^{x} = \vec{A}_{j} -  (\vec{A}_{j}\cdot \hat{\omega}_j) \ \hat{\omega}_j$, $\vec{A}_{j}^{y} =   \hat{\omega}_j\times \vec{A}_{j}$, and $I_{x}^j = \vec{I}_j \cdot \hat x_j$,  $I_{y}^j = \vec{I}_j \cdot \hat{y}_j$ with $\hat{x}_j = \vec{A}_{j}^{x}/ A_{j}^{x}$
and $\hat{y}_j = \vec{A}_{j}^{y}/ A_{j}^{y}$. 

Now, one selects a harmonic in the expansion of $F(t)$ and the period $T$, to create a resonant interaction of the NV with a target nucleus (namely the $k$th nucleus). To this end, in Eq.~(\ref{modulated}) we set $n=l$, and $T$ such that $l \omega_{\rm M} \approx \omega_k$. After eliminating fast rotating terms we get 
\begin{eqnarray}\label{big}
H &\approx& \frac{f_l A_k^x}{4}\sigma_z [I_k^{-} e^{i(\omega_k - l\omega_{\rm M}) t} +  {\rm H.c.}]\nonumber\\
&+&\sum_{j\neq k}  \frac{f_l A_j^x}{4}\sigma_z [I_j^{-} e^{i(\omega_j - l\omega_{\rm M}) t} +  {\rm H.c.}]\nonumber\\
&+&\sum_{n\neq l} \sum_j \frac{f_n A_j^x}{4}\sigma_z  [I_j^{-}e^{i(\omega_j - n\omega_{\rm M})t} + {\rm H.c.}].
\end{eqnarray}

By inspecting the first line of (\ref{big}), one finds that nuclear spin addressing at the $l$th harmonic is achieved when
\begin{equation}\label{resonancecond}
l \omega_{\rm M} = l  \frac{2\pi}{T}= \omega_k.
\end{equation}
With this resonance condition, the first line in (\ref{big}) is the resonant term $f_l A_k^x/4\sigma_z I_k^x$,  while detuned contributions (those in second and third lines) would average out by the rotating wave approximation (RWA). More specifically, with Eq.~(\ref{resonancecond}) at hand we can remove the second line in Eq.~(\ref{big}) if  
\begin{equation}\label{cond1}
|\omega_j - \omega_k| \gg  f_l A_j^x/4. 
\end{equation}
Detuned contributions corresponding to harmonics with $n\neq l$ are in the third line of~(\ref{big}). These can be neglected if
 \begin{equation}\label{cond2}
 |\omega_j - n/l \omega_k| \approx \omega_{\rm L} (l-n)/l \gg f_n A_j^x/4  \ \ \forall n. 
 \end{equation}
To strengthen condition~(\ref{cond1}), one can reduce the value of $f_l$ by selecting a large harmonic (see later), while condition~(\ref{cond2}) applies better for large values of $B_z$ since $\omega_{\rm L}\propto B_z$.

Assuming instantaneous $\pi$ pulses, standard DD sequences with  constant $\Omega$~\cite{Maudsley86, Gullion90, Souza11} lead to $|f_l| = \frac{4}{\pi l}, 0$ for $l$ odd, even. Thus, large harmonics (i.e. with large $l$) reinforce condition~(\ref{cond1}) as they lead to a smaller value for $f_l$. In Fig.~\ref{instantaneous} we compute the signal corresponding to the NV observable $\langle\sigma_x\rangle$ in a sample that contains 150 $^{13}$C nuclei ($\gamma_{^{13}{\rm C}} = (2\pi)\times 10.708$ MHz/T). To obtain sufficient spectral resolution we use large harmonics. Figure~\ref{instantaneous} (a) shows the signal for $l=13, 15$ and the theoretically expected values for $\langle\sigma_x\rangle$ (triangles for $l=13$ and circles for $l=15$) that would appear if perfect single nuclear addressing is considered~\cite{SupMat}. We observe that the computed signal does not match with the theoretically expected values. In addition to a flawed accomplishment of conditions~(\ref{cond1}, \ref{cond2}),  this is also a consequence of using large harmonics since, for large $l$, the period $T=2\pi l/\omega_k$ and the spacing between $\pi$ pulses grows, see  inset in Fig.~\ref{instantaneous}(a), spoiling the efficient elimination of the $\sigma_z A_{j}^{z} I^z_j$ terms in Eq.~(\ref{modulated}) by the RWA. In the inset of Fig.~\ref{instantaneous} (a) there is a sketch of the pulse structure we repeatedly apply ($20$ times in (a) and (b), while in (c) and (d) that structure is used 400 times) we  to get the signals in Fig.~\ref{instantaneous}, red blocks are instantaneous $\pi$ pulses, while their associated $F(t)$ is in blue. Working with even larger harmonics introduces the problem of spectral overlaps. These appear when the signal associated to a certain harmonic contains resonance peaks corresponding to other harmonics. In Fig.~\ref{instantaneous} (b) one can see (green arrow) how a peak of $l=35$ (green circle) is mixed with the signal of $l=33$ (orange triangle). This is an additional disadvantage since the interpretation of the spectrum gets challenging.

Condition~(\ref{cond2}) is strengthened using a large $B_z$ as $\omega_{\rm L} \propto B_z$. This also implies a larger resonance frequency (namely $\omega_k$) for each nucleus. Addressing large $\omega_k$ is beneficial since the period $T$ (note that, in resonance $T=2\pi l/\omega_k$) and the interpulse spacing get shorter turning into a better cancellation of $\sigma_z A_{j}^{z} I^z_j$ terms. In Fig.~(\ref{instantaneous}) (c) (d), we use a large $B_z=1$ T and the spectral overlap is removed, while the computed signal matches the theoretically expected values (blue triangles). 

Unfortunately, to consider $\pi$ pulses as instantaneous in situations with large $B_z$ is not correct, since nuclei have time to evolve during $\pi$ pulse execution leading to signal drop~\cite{Casanova18}.  Hence, if one cannot deliver a huge MW power to the sample, the results in Fig.~\ref{instantaneous} (c) and (d) are not achievable.

\section{A solution with extended pulses}
\begin{figure*}[t]
\hspace{-0.25 cm}\includegraphics[width=1.9\columnwidth]{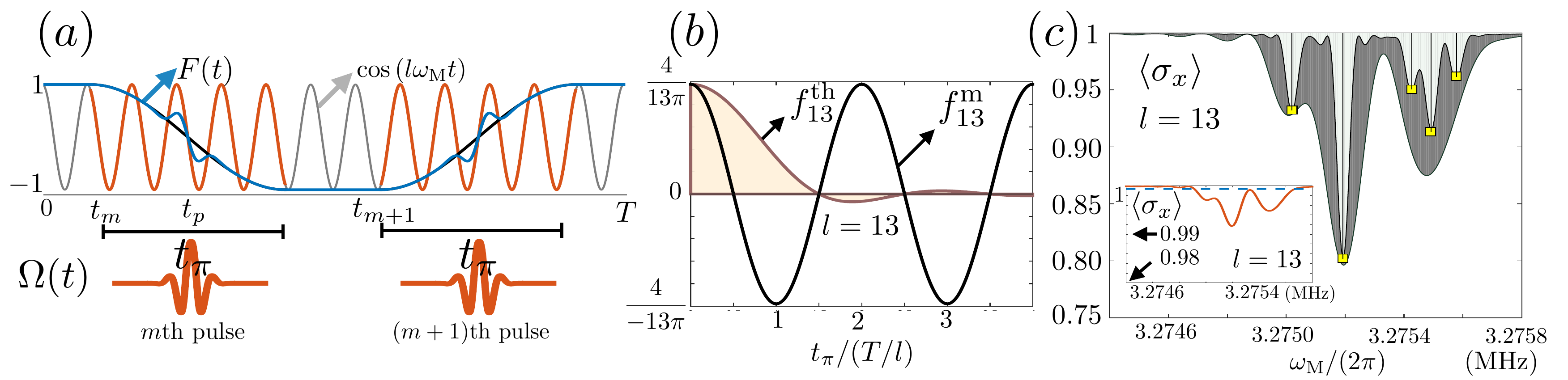}
\caption{(a) Upper panel, one period of $F(t)$ (solid-blue) including the intrapulse behavior, and the $\cos{(l\omega_{\rm M}t)}$ function. Extended $\pi$ pulses span during $t_{\pi}$ (intrapulse regions appear marked in red). In this example, $l=13$ and $t_{\pi}\approx4.5\times(T/l)$. Solid-black, behavior of $F(t)$ in case standard top-hat pulses are applied. Bottom panel, train of modulated $\Omega(t)$ leading to $F(t)$. (b) $f_l^{\rm m}$ (black-solid) and $f_l^{\rm th}$ (curve on the yellow area) as a function of the ratio $t_{\pi}/(T/l)$ for $l=13$. (c) $\langle \sigma_x\rangle$ (curves over dark and clear areas) for the conditions discussed in the main text. Inset, $\langle \sigma_x\rangle$ computed with top-hat pulses. For all numerical simulations in (c) we assume a $1\%$ of error in  $\Omega(t)$~\cite{Cai12}.}
\label{whole}
\end{figure*}

In realistic situations $\pi$ pulses are finite, thus the value of $f_l=2/T\int_0^T F(s)  \cos{(l \omega_{\rm M} s)} ds$ has to be computed by considering the intrapulse contribution. This is (for a generic $m$th pulse)  $2/T  \int_{t_m}^{t_m+t_{\pi}} F(s) \cos{(l\omega_{\rm M} s)} \ ds$, with $t_\pi$ being the $\pi$ pulse time and $t_m$ the instant we start applying MW radiation, see Fig.~\ref{whole} (a). In addition, the  $F(t)$ function must hold the following conditions: Outside  the $\pi$ pulse region $F(t)=\pm 1$, while $F(t)$ is bounded as $-1 \leq F(t) \leq 1$ $\forall t$, Fig.~\ref{whole} (a). 

Now, we present a design for $F(t)$ that satisfies the above conditions, cancels intrapulse contributions, and leads to tunable NV-nuclei interactions. In particular, for the $m$th pulse
\begin{equation}\label{modulatedF}
F(t) = \cos{\big[\pi(t - t_{m})/t_\pi \big]} + \sum_{q}\alpha_{q}(t) \sin{\big[ q l \omega_{\rm M}(t - t_{p}) \big]}.
\end{equation} 
Here, $\alpha_{q}(s)$ are functions to be adjusted (see later) and $t_p=t_m+t_{\pi}/2$ is the central point of the $m$th pulse, Fig.~\ref{whole} (a). We modulate $F(s)$ in the intrapulse region such that  (for the $m$th pulse) $\int_{t_m}^{t_m+t_{\pi}} F(s) \cos{(l\omega_{\rm M} s)} \ ds =0$, this is $F(t)$ cancels the intrapulse contribution. Once we have $F(t)$, we find the associated Rabi frequency $\Omega(t)$ with the formula $\Omega(t) = \frac{\partial}{\partial t} \arccos[F(t)]$~\cite{SupMat}. Now, the value of the $f_l$ coefficient obtained with the {\it modulated} $F(t)$ in Eq.~(\ref{modulatedF}) (from now on denoted $f_l^{\rm m}$) depends only on the integral out of $\pi$ pulse regions. This can be calculated leading to~\cite{SupMat}
\begin{equation}\label{modulatedf}
f_l^{\rm m} = \frac{4}{\pi l }\cos{\bigg(\pi  \frac{t_{\pi}}{T/l}\bigg)}\sin{(\pi l /2)},
\end{equation} 
which is our main result. By modifying the ratio between $t_\pi$ (the extended $\pi$ pulse length) and $T/l$ we can select a value for $f^{\rm m}_l$ and achieve tunable NV-nuclei interactions. According to Eq.~(\ref{modulatedf}), $f_l^{\rm m}$ can be taken to any amount between $-\frac{4}{l\pi}$ and $\frac{4}{l\pi}$, see solid-black curve in Fig.~\ref{whole} (b). In addition, owing to the periodic character of Eq.~(\ref{modulatedf}), one can get an arbitrary value (between $-\frac{4}{l\pi}$ and $\frac{4}{l\pi}$) for $f_l^{\rm m}$ even with large  $t_{\pi}$.  This implies highly extended $\pi$ pulses, thus a low delivered MW power. On the contrary, for standard $\pi$ pulses in the form of {\it top-hat} functions (i.e. generated with constant $\Omega$) one finds~\cite{SupMat} 
\begin{equation}
f_l^{\rm th} = \frac{4\sin{(\pi l /2)}\cos{(\pi l t_{\pi}/T)}}{\pi l (1-4l^2t_{\pi}^2/T^2 )}.
\end{equation}
Unlike $f^{\rm m}_l$, the expression for $f_l^{\rm th}$ shows a decreasing fashion for growing  $t_{\pi}$. Note that $ |f_l^{\rm th}| \propto [t_{\pi}/(T/l)]^{-2}$. This behaviour can be observed in Fig.~\ref{whole} (b), curve over the yellow area. Hence, standard top-hat pulses cannot operate with a large $t_{\pi}$, as this leads to a strong decrease of $f_l^{\rm th}$, thus to signal loss. 

To show the performance of our theory, we select, a gaussian form for $\alpha_{1}(t) = a_1 e^{-(t-t_p)^2/2c^2}$  and set $\alpha_{q}(t)=0$, $\forall q>1$. See one example of a modulated $F(t)$ in Fig.~(\ref{whole}) (a) (solid-blue) as well as the behavior of $F(t)$ if common top-hat $\pi$ pulses are used (solid-black). Once we choose the $t_{\pi}$, $l$, and $c$ parameters that will define the shape of $F(t)$, we select the remaining constant $a_1$ such that it  cancels the intrapulse contribution, i.e. $\int_{t_m}^{t_m+t_{\pi}} F(s) \cos{(l\omega_{\rm M} s)} \ ds =0$. By inspecting Eq.~(\ref{modulatedF}) one easily finds that a natural fashion for $a_1$ is given by
\begin{equation}\label{a1ratio}
a_1=-\frac{\int_{t_m}^{t_m+t_\pi} \cos{\big[\pi(s - t_{m})/t_\pi \big]}  \cos{(l\omega_{\rm M} s)} \ ds}{\int_{t_m}^{t_m+t_\pi} e^{-\frac{(s-t_p)^2}{2c^2}}\sin{\big[ l \omega_{\rm M}(s - t_{p}) \big]}  \cos{(l\omega_{\rm M} s)} \ ds}.
\end{equation}

In Fig.~\ref{whole} (c) we simulated a sample containing 5 protons~\cite{Note} at an average distance from the NV of $\approx 2.46$ nm.  Numerical simulations have been performed starting from Eq.~(\ref{simulations}) without doing further assumptions. The 5-H target cluster has the hyperfine vectors (note $\gamma_{\rm H} = (2\pi)\times 42.577$ MHz/T) $\vec{A}_1 = (2\pi)\times[-1.84, -3.19, -11.02]$, $\vec{A}_2 = (2\pi)\times[2.38, 5.04, -8.78]$, $\vec{A}_3 = (2\pi)\times[8.09, 2.66, -1.02]$,  $\vec{A}_4 = (2\pi)\times[4.26, 2.46, 3.48]$, and $\vec{A}_5 = (2\pi)\times[4.07, 1.00, -7.09]$ kHz. We simulate two different sequences, leading to two signals, using our extended $\pi$ pulses  under a large magnetic field $B_z=1$ T. Vertical panels with yellow squares mark the theoretically expected resonance positions and signal contrast. For the first computed signal, curve over dark area in Fig.~\ref{whole} (c), we display a XY-8 sequence where each X (Y) extended pulse has $\phi=0$ ($\phi=\pi/2$). The modulated Rabi frequency $\Omega(t)$ in $H'_{\rm c}$ is selected  such that it leads to $f_{13}^{\rm m}=4\pi/13 = 0.0979$ for $l=13$ (note this corresponds to the maximum value for $f_{13}^{\rm m}$) with  a pulse length $t_{\pi} = 6 \times(T/l)$. In addition, we take the width of the Gaussian function $\alpha_{1}(t)$ as $c = 0.07 t_{\pi}$. The scanning frequency $\omega_{\rm M}$ spans around $\gamma_{\rm H} B_{z}/l$ for $l=13$, see horizontal axis in Fig.~\ref{whole} (c).  After repeating the XY-8 sequence  $400$ times, i.e. 3200 extended  $\pi$ pulses have been applied leading to a final sequence time of $t_{f} \approx 0.488$ ms, we get the signal over the dark area.  As we observe in Fig.~\ref{whole} (c), this sequence does not resolve all nuclear resonances of the 5-H cluster. 

To overcome this situation, we make use of the tunability of our method, and simulate a second sequence with extended $\pi$ pulses  leading to the signal over the clear area in Fig.~\ref{whole}~(c). This has been computed with a smaller value for $f_{13}^{\rm m} = 0.0979/3 = 0.0326$ which is achieved with $t_{\pi} \approx 6.4\times(T/l)$, i.e. a slightly longer $\pi$ pulse than those in the preceding situation, and $c=0.07 t_{\pi}$.  As the $f_{13}^{\rm m}$ coefficient is now smaller, we have repeated the XY-8 sequence $400\times 3$ times (i.e. 9600 pulses) to get the same contrast than in the previous case. The final time of the sequence is $t_{f} \approx 1.5$ ms. As we observe in Fig.~\ref{whole} (c), our method faithfully resolves all resonances in the 5-H cluster, and reproduces the theoretically expected signal contrast. It is noteworthy to comment that the tunability offered by our method will be of help for different quantum algorithms with NV centers~\cite{Ajoy15, Perlin18, Casanova16, Casanova17}.

\section{Microwave power and nuclear signal comparison}
In the inset of Fig.~\ref{whole} (c) we plot the signals one would get using standard top-hat pulses with the same average power than our extended pulses in  Fig.~\ref{whole} (c).  We use that the energy of each top-hat and extended $\pi$ pulse, $E^{top-hat}(t_\pi)$ and $E^{extended}(t_\pi)$, is $\propto \int \Omega^2(s) ds$ where the integral extends during the $\pi$ pulse duration (top-hat or extended). For an explicit  derivation of the energy relations see~\cite{SupMat}. The solid-orange signal in the inset has been computed with a XY-8 sequence containing  3200 top-hat $\pi$ pulses with a constant $\Omega \approx (2\pi)\times 18.2$ MHz. For this value of $\Omega$, a top-hat $\pi$ pulse contains the same average power than each extended $\pi$ pulse used to compute the signal over dark area in  Fig.~\ref{whole} (c), i.e. $E^{top-hat}(t_\pi)= E^{extended}(t_\pi)$. Unlike our method, the sequence with standard top-hat $\pi$ pulses produces a signal with almost no-contrast. Note that the vertical axis of inset in Fig.~\ref{whole} (c) has a maximum depth value of 0.98, and the highest contrast achieved with top-hat pulses falls below 0.99. The dashed signal in the  inset has been obtained with top-hat $\pi$ pulses with $\Omega\approx(2\pi)\times 4.68$ MHz. Again, this is done to assure we use the same average power than the sequence leading to the curve over the clear area in Fig.~\ref{whole} (c). In this last case, we observe that the signal harvested with standard top-hat $\pi$ pulses does not show any appreciable contrast. These results indicate that our method using pulses with modulated amplitude is able to achieve tunable electron nuclear interactions, while regular top-hat pulses with equivalent MW power fail to resolve these interactions.

\section{Conclusion}
We presented a general method to design extended $\pi$ pulses which are energetically efficient, and incorporable to stroboscopic DD techniques such as the widely used XY-8 sequence. Our method leads to tunable interactions, hence selective, among an NV quantum sensor and nuclear spins at large static magnetic fields which represents optimal conditions for nanoscale NMR.

\begin{acknowledgements}
The authors thank J. F. Haase for commenting on the manuscript. Authors acknowledge financial support from Spanish MINECO/FEDER FIS2015-69983-P and PGC2018-095113-B-I00 (MCIU/AEI/FEDER, UE), Basque Government IT986-16, as well as from QMiCS (820505) and OpenSuperQ (820363) of the EU Flagship on Quantum Technologies. J.C. acknowledges support by the Juan de la Cierva grant IJCI-2016-29681. I. A. acknowledges support to the Basque Government PhD grant PRE-2015-1-0394. This material is also based upon work supported by the U.S. Department of Energy, Office of Science, Office of Advance Scientific Computing Research (ASCR), Quantum Algorithms Teams project under field work proposal ERKJ335.
\end{acknowledgements}

\pagebreak
\widetext
\begin{center}
\textbf{ \large Supplemental Material: \\ Selective Hybrid Spin Interactions with Low Radiation Power}
\end{center}

\setcounter{equation}{0} \setcounter{figure}{0} \setcounter{table}{0}
\setcounter{page}{1} \makeatletter \global\long\def\theequation{S\arabic{equation}}
 \global\long\def\thefigure{S\arabic{figure}}
 \global\long\def\bibnumfmt#1{[S#1]}
 \global\long\def\citenumfont#1{S#1}

\section{Ideal signal under single nuclear addressing}
In case of having perfect single nuclear addressing with the $k$th nucleus and the $l$th harmonic, Eq.~(5) in the main text can be reduced to
\begin{equation}
H= \frac{f_lA_k^x}{4}\sigma_z I_k^x.
\end{equation}
For the above Hamiltonian the dynamics can be exactly solved, and the evolution of $\langle \sigma_x \rangle$ (when the initial state is $\rho=|+\rangle\langle +| \otimes \frac{1}{2} \mathbb{I}$, i.e. we consider the nucleus in a thermal state) reads
\begin{equation}
\langle \sigma_x \rangle = \cos{\bigg(\frac{f_lA_k^{x}}{4}t\bigg)}.
\end{equation} 
The above expression represents the depth of each panel (circles or triangles) in Fig. 1. of the main text.

\section{Finding $\Omega(t)$ from $F(t)$}
The MW driving in Eq.~(2) is $\frac{\Omega(t)}{2} (|1\rangle\langle 0| e^{i\phi} +{\rm H.c.})$, and its propagator for, e.g., the $m$th $\pi$-pulse is $U_t = e^{-i\int_{t_m}^{t_m+t_{\pi}} \frac{\Omega(s)}{2} (|1\rangle\langle 0| e^{i\phi} +{\rm H.c.}) \ ds}$. During the $m$th $\pi$-pulse, i.e. in a certain time between $t_m$ and $t_m + t_{\pi}$, $U_t $  has the following effect on the electron spin $\sigma_z$ operator (in the following we call $\sigma_\phi = |1\rangle\langle 0| e^{i\phi} +{\rm H.c.}$)
\begin{equation}
e^{i\int_{t_m}^{t_m+t} \frac{\Omega(s)}{2} \sigma_\phi \ ds} \sigma_z e^{-i\int_{t_m}^{t_m+t} \frac{\Omega(s)}{2} \sigma_\phi \ ds} = e^{\big(i\int_{t_m}^{t_m+t} \Omega(s) \ ds \big)  \sigma_\phi }  \sigma_z = \cos{\bigg(\int_{t_m}^{t_m+t} \Omega(s) \ ds \bigg)} \sigma_z + i  \sin{\bigg(\int_{t_m}^{t_m+t} \Omega(s) \ ds \bigg)} \sigma_{\phi}\sigma_z.
\end{equation}
In this manner, we can say that $F(t) = \cos{\big(\int_{t_m}^{t_m+t} \Omega(s) \ ds \big)}$ while the other spin component, i.e. the one going with $ \sin{\big(\int_{t_m}^{t_m+t} \Omega(s) \ ds \big)}$, does not participate in the joint NV-nucleus dynamics for sequences with alternating pulses~\cite{Lang17} such as the XY-8$\equiv$ XYXYYXYX pulse sequence we are using in the article. Now, one can easily invert the expression $F(t) = \cos{\big(\int_{t_m}^{t_m+t} \Omega(s) \ ds \big)}$ and find $\Omega(t) = \frac{\partial}{\partial_t} \arccos[F(t)]$. The latter is the expression mentioned in the main text.

\section{Calculation of $f_l$ coefficients}
\subsection{Coefficients for extended pulses}

The analytical expression for the coefficients $f_l$ is given by 
\begin{equation}
f_{l}=\frac{2}{T}\int_{0}^{T} F(s) \cos{\Big(\frac{2 \pi l s}{T}\Big)}  \ ds,
\end{equation}
where $T=2 \pi /\omega_{\rm M}$. With a rescaling of the integrating variable given by $s=xT/2 $, this is rewritten as  
\begin{equation}\label{equation22}
f_{l}=\int_{0}^{2} F(x) \cos{(\pi l x)}  \ dx.
\end{equation}
The function inside the integral is symmetric or antisimmetric w.r.t. $x=1$, depending on $l$ been odd or even. This can be easily demonstrated by using $F(x+1)=-F(x)$ and $\cos{(\pi l (x+1))}=\cos(\pi l)\cos{\pi l x}$. Thus, if $l$ is even and the function is symmetric w.r.t $x=1$, the value of the integral will be zero. Anyway, one can work a general expression for Eq.(\ref{equation22}). First, we can divide the integral in two parts,
\begin{equation}
f_{l}=\int_{0}^{1} F(x) \cos{(\pi l x)}  \ dx +\int_{1}^{2} F(x) \cos{(\pi l x)}  \ dx,
\end{equation}
and substitute $x$ for $x+1$ in the second integral. Using the symmetry properties specified above, the equation reduces to 
\begin{equation}
f_{l}=(1-\cos{(\pi l)})\int_{0}^{1} F(x) \cos{(\pi l x)}  \ dx.
\end{equation}
Now, from $x=0$ to $1$, $F(x)$ can be divided in three parts defined by $\tau_m\equiv 2 t_{m}/T$ and $1-\tau_m$,
\begin{equation}\label{integrals}
f_{l}=(1-\cos{(\pi l)})\Bigg\{\int_{0}^{\tau_m} F(x) \cos{(\pi l x)}  \ dx +\int_{\tau_m}^{1-\tau_m} F(x) \cos{(\pi l x)}  \ dx+\int_{1-\tau_m}^{1} F(x) \cos{(\pi l x)}  \ dx\Bigg\},
\end{equation}
and the integral in the middle is zero for the extended pulses. This leaves us with the first and third integrals for which $F(x)$ is $1$ and $-1$ respectively, obtaining 
 \begin{equation}
f_{l}^{\rm m}=(1-\cos{(\pi l)})\Bigg\{\int_{0}^{\tau_m} \cos{(\pi l x)}  \ dx -\int_{1-\tau_m}^{1}  \cos{(\pi l x)}  \ dx\Bigg\},
\end{equation}
that leads to
 \begin{equation}
f_{l}^{\rm m}=\frac{1}{\pi l }(1-\cos{(\pi l)})\Bigg\{\sin{(\pi l \tau_m)} + \sin{(\pi l (1-\tau_m))} \Bigg\}.
\end{equation}
Using $\sin{(\pi l (1-\tau_m))}=-\sin{(\pi l \tau_m)}\cos(\pi l)$ and $\sin^2{\theta}=(1-\cos{(2\theta)})/2$, the expression for $f_{l}^{\rm m}$ reduces to 
 \begin{equation}
f_{l}^{\rm m}=\frac{4}{\pi l }\sin^4{(\pi l/2)}\sin{(\pi l \tau_m)}.
\end{equation}
Now, by using the relation $T=4t_m+2t_\pi$,  $f_{l}^{\rm m}$ becomes
\begin{equation}\label{eqbat}
f_{l}^{\rm m}=\frac{4}{\pi l }\sin^4{(\pi l/2)}\sin{\Big(\pi l \Big(\frac{1}{2}+\frac{t_\pi}{T}\Big)\Big)},
\end{equation}
where $t_{\pi}$ is the duration of a $\pi$-pulse. Eq.(\ref{eqbat}) is equivalent to Eq.(10) in the main text. To prove that, one may use the trigonometric identity $\sin(\theta+\pi l /2)=\sin{(\theta)}\cos{(\pi l /2)} +\cos{(\theta)}\sin{(\pi l /2)}$ which leads us to 
\begin{equation}\label{eqlast}
f_{l}^{\rm m}=\frac{4}{\pi l }\cos{\Bigg(\pi  \frac{t_{\pi}}{T/l}\Bigg)}\sin{(\pi l /2)},
\end{equation}
as $\sin^4{(\pi l /2)}\cos{(\pi l /2)}=0$ and $\sin^5{(\pi l /2)}=\sin{(\pi l /2)}$. 

\begin{figure}[t]
\includegraphics[width=0.9\columnwidth]{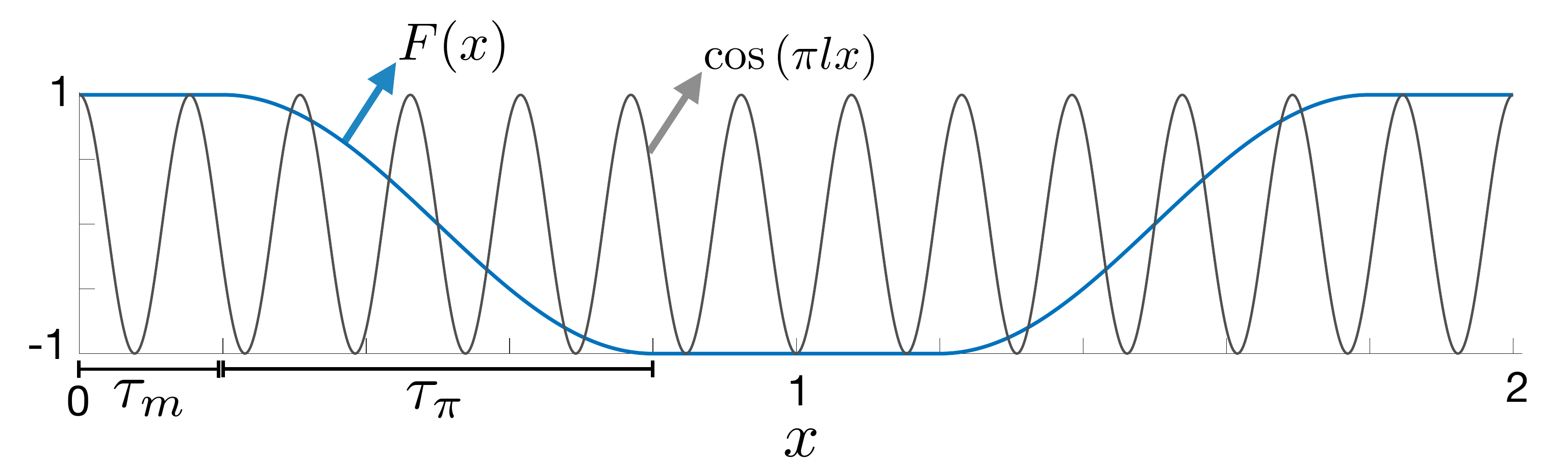}
\caption{Plot of $F(x)$ and $\cos{(\pi l x)}$ (where $l=13$) functions between $x=0$ and $x=2$, corresponding to $t=0$ and $t=T$ respectively. }
\label{Subfig}
\end{figure}

\subsection{Coefficients for top-hat pulses}
For calculating the value of $f_l$ coefficients in the case of top-hat pulses, we just need to sum the contribution of the second integral on Eq.(\ref{integrals}), which is not zero for top-hat pulses. The value of $F(s)$ during the pulse can be written as 
\begin{equation}
F(s)=\cos{[\pi(s-t_m)/t_\pi]},
\end{equation}
where $t_{p}=t_m+t_\pi/2$. With the rescaling of the integrating variable introduced in the previous section this is rewritten as
\begin{equation}
F(s)=\cos{[\pi(x-\tau_m)/\tau_\pi]},
\end{equation}
where $\tau_\pi=2 t_\pi /T $. So, we need to solve the following integral 
\begin{equation}
\int_{\tau_m}^{1-\tau_m} F(x) \cos{(\pi l x)}  \ dx=\int_{\tau_m}^{1-\tau_m}  \cos{[\pi(x-\tau_m)/\tau_\pi]}\cos{(\pi l x)}  \ dx
\end{equation}
which is not zero. To solve the integral, we can displace the reference frame by a factor of $\tau_p=1/2$, by the change of variable $x=y+\tau_m+\tau_\pi/2=y+1/2$. Now, the integral will be centered at zero and will look like 
\begin{equation}
\int_{-\tau_{\pi}/2}^{\tau_{\pi}/2}  \cos{[\pi y/\tau_{\pi}+\pi/2]} \cos{[\pi l (y+1/2)]}  \ dy=-\int_{-\tau_{\pi}/2}^{\tau_{\pi}/2}  \sin{(\pi y/\tau_{\pi})} \cos{[\pi l (y+1/2)]}  \ dy,
\end{equation}
and using $\cos[\pi l (y+1/2)]=\cos{(\pi l y)}\cos{(\pi l /2)} - \sin{(\pi l y)}\sin{(\pi l /2)}$ becomes
\begin{equation}
\sin{(\pi l /2)}\int_{-\tau_{\pi}/2}^{\tau_{\pi}/2}  \sin{(\pi y/\tau_{\pi})} \sin{(\pi l y)}  \ dy - \cos{(\pi l /2)}\int_{-\tau_{\pi}/2}^{\tau_{\pi}/2}  \sin{(\pi y/\tau_{\pi})} \cos{(\pi l y)}  \ dy
\end{equation}
where the second integral is zero owing to symmetry reasons, i. e. $\int_{-a}^{a}F(x)dx=0$ if $F(-x)=-F(x)$. Again, because of symmetry arguments, the first integral is
\begin{equation}
2\sin{(\pi l /2)}\int_{0}^{\tau_{\pi}/2}  \sin{(\pi y/\tau_{\pi})} \sin{(\pi l y)}  \ dy,
\end{equation}
which using trigonometric identities reads 
\begin{equation}
\sin{(\pi l /2)}\Big\{\int_{0}^{\tau_{\pi}/2}  \cos{(\pi y(l-1/\tau_{\pi})}   \ dy -\int_{0}^{\tau_{\pi}/2}  \cos{(\pi y(l+1/\tau_{\pi})}   \ dy.\Big\}
\end{equation}
Solving the integrals one gets
\begin{equation}
\frac{-1}{\pi}\sin{(\pi l /2)}\cos{(\pi l\tau_{\pi}/2)}\Big\{ \frac{1}{l-1/\tau_{\pi}}  +  \frac{1}{l+1/\tau_{\pi}} \Big\},
\end{equation}
which is simplified to 
\begin{equation}
\frac{2l\tau_\pi^2}{\pi(1-l^2\tau_{\pi}^2)}\sin{(\pi l /2)}\cos{(\pi l\tau_{\pi}/2)}.
\end{equation}
It is straightforward to prove that the sum of  the three integrals in Eq.~(\ref{integrals}) gives 
\begin{equation}
f_{l}^{\rm th}=\frac{4 \sin{(\pi l /2)}\cos{(\pi l t_{\pi}/T)}}{\pi l(1-4l^2t_{\pi}^2/T^2)},
\end{equation}
which correspond to the expression written in the main text.

\section{Energy delivery}
The Poynting vector, that describes the energy flux for an electromagnetic wave, is given by 
\begin{equation}
\vec{P}=\frac{1}{\mu_{0}} \vec{E} \times \vec{B},
\end{equation}
where $\mu_{0}$ is the vacuum permeability, and $\vec{E}$ and $\vec{B}$ are the electric field and magnetic field vectors at the region of interest, i.e. the NV center. The latter, in the nanoscale, is sufficiently small compared with the wavelength of the microwave (MW) radiation to assume a plane wave description of the radiation, so the magnetic field can be written as
\begin{equation}
\vec{B}=\vec{B}_{0}(t)\cos{(\vec{k}\cdot \vec{x}-\omega t +\varphi)},
\end{equation}
where $\vec{k}$ is the wavevector and $\omega$ the frequency of the microwave field. We will also assume an extra time dependence $B_{0}(t)$ whose time scales will be several orders of magnitude larger than the period $2\pi/\omega$. From Maxwell equations in vacuum it is derived that, for such a magnetic field, $\vec{k}\cdot \vec{B}=0$, $\vec{k}\cdot \vec{E}=0$, and $\vec{E}\cdot \vec{B}=0$. From the equation $\vec{\nabla}\times\vec{B}=\frac{1}{c^2}\partial \vec{E}/\partial t$, it follows that
\begin{equation}\label{electric}
\vec{E}=c^2\int\!dt \ (\vec{\nabla}\times\vec{B})=-c^2\int\! dt \ (\vec{k}\times\vec{B_0}(t)) \sin{(\vec{k}\cdot \vec{x}-\omega t +\varphi)}.
\end{equation}
We choose $\vec{B}$ to be perpendicular to the NV axis ($z$ axis), specifically, on the $x$ axis. The control Hamiltonian, is then 
\begin{equation}
H_{c}(t)=-\gamma_e\vec{B}\cdot \vec{S}=\gamma_eB_{x}(t)S_x \cos{(\omega t - \phi)},
\end{equation}
where $\vec{S}$ corresponds to the spin of the NV center, $\gamma_e$ is the gyromagnetic ratio of the electron and $\vec{x}=0$ the position of the NV. To recover Eq.(1) of the main text, we require that $\sqrt{2} \Omega(t)=\gamma_e B_x(t)$. The magnetic field vector at $\vec{x}=0$ is then
\begin{equation}
\vec{B}(t)=\frac{\sqrt{2}\Omega(t)}{\gamma_e} \cos{(\omega t -\varphi)} \hat{x}\\
\end{equation}
and the electric field is, from Eq.(\ref{electric}),
\begin{equation}
\vec{E}(t)=\frac{\sqrt{2}\omega c}{\gamma_e}\int dt \Omega(t) \sin{(\omega t -\varphi)} \ \hat{k}\times \hat{x},
\end{equation}
which, using the wave equation $\partial^2 \vec{E}/\partial^2 t=c^2\nabla^2 \vec{E}=\omega^2\vec{E}$, converts into
\begin{equation}\label{elec2}
\vec{E}(t)=\frac{\sqrt{2}}{k\gamma_e} \frac{\partial}{\partial t}\Big[ \Omega(t) \sin{(\omega t -\varphi)} \Big]\ \hat{x}\times \hat{k}.
\end{equation}

\subsection{The case of top-hat $\pi$ pulses}

For top-hat pulses we have that $\partial\Omega(t)/\partial t=0$, thus, the energy delivery per unit of area we obtain for top-hat pulses is
\begin{equation}
E^{\rm top-hat}(t_{\pi})=\int_0^{t_\pi}dt |\vec{P}(t)|=\frac{c}{\mu_0}\frac{2}{\gamma_{e}^2}\int_0^{t_\pi}dt \ \Omega^2\cos^2(\omega t-\varphi)=\frac{c}{\mu_0}\frac{\Omega^2}{c\gamma_e^2}\int_{0}^{t_\pi} dt\Big\{1+\cos{(2\omega t -2\varphi)}\Big\},
\end{equation}
which gives
\begin{equation}\label{eth}
E^{\rm top-hat}(t_{\pi})=\frac{c}{\mu_0}\frac{\Omega^2}{\gamma_e^2} \Big\{t_\pi+\frac{1}{2\omega}\sin{(2\omega t_\pi -2\varphi)}\Big\}.
\end{equation}
The second part of the formula is upper bounded by $(2\omega)^{-1}$, which , on the other hand, is several orders of magnitude smaller than $t_\pi$, thus negligible. As $t_{\pi}=\pi/\Omega$, Eq.(\ref{eth}) can be rewritten as
\begin{equation}\label{Stophat}
E^{\rm top-hat}(t_{\pi})\approx \frac{\pi c}{\mu_0}\frac{ \Omega}{\gamma_e^2},
\end{equation}
meaning that the energy increases linearly with the Rabi frequency.

\subsection{The case of extended $\pi$ pulses}
To study the case of an extended $\pi$ pulse, we need to calculate both terms on Eq.(\ref{elec2}), which are non zero in general. The complete expression is given by 

\begin{equation}\label{Sextended}
E^{\rm extended}(t_{\pi})=\frac{c}{\mu_0}\frac{2}{\gamma_e^2}\int_{0}^{t_\pi}\! \!dt \ \Bigg[ \Omega^2(t)\cos^2{(\omega t -\varphi)} +\frac{1}{\omega} \Omega(t)\frac{\partial\Omega(t)}{\partial t}\cos{(\omega t -\varphi)}  \sin{(\omega t -\varphi)} \Bigg] .
\end{equation}
As a final comment, for all cases simulated in the main text we find that the second term at the right hand side of Eq.~(\ref{Sextended}) is negligible, thus it can be written 
\begin{equation}
E^{\rm extended}(t_{\pi})\approx \frac{c}{\mu_0}\frac{2}{\gamma_e^2}\int_{0}^{t_\pi}\! \!dt \ \Bigg[ \Omega^2(t)\cos^2{(\omega t -\varphi)}  \Bigg] .
\end{equation}

\subsection{Equivalent top-hat Rabi frequency}
To calculate the constant Rabi frequency leading to top-hat pulses with the same energy than extended pulses, one has to equal $E^{\rm top-hat}(t_{\pi}) = E^{\rm extended}(t_{\pi})$ and extract the value of the constant $\Omega$. With Eqs.~(\ref{Stophat}, \ref{Sextended}) one can easily find that 
\begin{equation}
\Omega=\frac{\mu_0 \gamma_e^2}{\pi c} E^{\rm extended}(t_{\pi}).
\end{equation}


\begin{thebibliography}{99}
\bibitem{SupMat} See Suplemental Material at 
[URL will be inserted by publisher for [further information on the calculations.]
\bibitem{Note} Note that the presence of extended pulses disables the use of numerical techniques, as those in~\cite{Maze08}, to simulate large nuclear samples. Hence, because of machine restrictions, here we focus in a sample that includes 5 nuclei and the NV.
\end{thebibliography}
\end{document}